\documentclass[twocolumn]{aa}
\usepackage{graphicx,epsf,epsfig}
\usepackage{txfonts}
\newcommand{\rx}{RX\,J2133.7+5107}
\newcommand{\Ha}{H$_{\alpha}{ }$}
\newcommand{\Hb}{H$_{\beta} $ }
\newcommand{\Hg}{H$_{\gamma}{ }$ }

\newcommand{\HeII} {HeII(~4686\,\AA) }
\newcommand{\HeIIb} {HeII(~5411\,\AA) }

\newcommand{\Mwd}{\mbox{$M_{\rm wd}$}}

\newcommand{\Msun}{\mbox{$M_{\odot}\,$}}

\newcommand{\ecsa}{\mbox{$\rm ergs\;cm^{-2}s^{-1}\mbox{\AA}^{-1}$}}
\newcommand{\ecs}{\mbox{$\rm ergs\;cm^{-2}s^{-1}$}}
\newcommand{\es}{\mbox{$\rm ergs\;s^{-1}$}}

\newcommand{\kms}{\mbox{$\rm km\,s^{-1}$}}
\newcommand{\draft}[1]{
}
\draft{Draft 1.0, \today}
\begin{document}

\title{RX J2133.7+5107 : Identification of a new long period Intermediate Polar
\thanks{Based on observations obtained at the Haute-Provence Observatory (France) and
at the Loiano Observatory (Italy), operated by the Istituto Nazionale di Astrofisica}
\thanks{Tables 2 and 3 and Figures 4-5-6 are only available in electronic form at 
http://www.edpsciences.org}
}
                                          
\titlerunning{RX J2133.7+5107: a new long period Intermediate Polar}

\author{J.M. Bonnet-Bidaud\inst{1}
\and M. Mouchet\inst{2,3}
\and D. de Martino\inst{4}
\and R. Silvotti\inst{4}
\and C. Motch\inst{5}
         }

\offprints{J.M. Bonnet-Bidaud, email: bonnetbidaud@cea.fr}

\institute{
Service d'Astrophysique, DSM/DAPNIA/SAp, CE Saclay, F-91191 Gif sur Yvette
Cedex, France\\
\email{bonnetbidaud@cea.fr}
\and
LUTH(CNRS-UMR8102), Observatoire de Paris, Section de Meudon, F-92195 Meudon Cedex, France
\and
APC (CNRS-UMR7164), Universit\'e Denis Diderot, 2 Place Jussieu, F-75005 Paris, France\\
\email{martine.mouchet@obspm.fr} 
\and
INAF--Osservatorio Astronomico di Capodimonte, Via Moiariello 16, I-80131 Napoli, Italy\\
\email{demartino@na.astro.it}
\and
Observatoire Astronomique de Strasbourg, F-67000 Strasbourg, France\\
 \email{motch@astro.u-strasbg.fr}
             }

   \date{Received: xx, xxxx; accepted: xx, xxxx}

   \abstract{
We report the first time-resolved photometric and spectroscopic optical observations of the 
X-ray source \rx, identified in the ROSAT survey. 
A clear persistent optical light pulsation is discovered with fast photometry 
at a period 
of $\rm P_{\omega}$=(570.823$\pm$0.013)\,s which we associate 
with the spin period of an accreting white dwarf. 
Radial velocity curves of the strong emission lines show modulation with 
a period of $\rm P_{\Omega}$=(7.193$\pm$0.016)\,hr, identified as the orbital period.  
These observations establish that the source is a member of the intermediate 
polar class (IPs) of magnetic cataclysmic variables. With only 4 IPs with longer 
orbital periods, \rx\, is among the widest systems. It is a unique IP
with an orbital period in the middle of the 
so-called (6-10)hr IP gap and it shows a significant degree 
of asynchronism with a ratio $\rm P_{\omega}$/$\rm P_{\Omega}$ of 0.02. 
When attributed to the motion of the white dwarf, the emission lines orbital 
modulation yields a mass function of 
f$_m$ = (1.05$\pm$0.21) $\times 10^{-2}$ \Msun which, for a probable 
inclination $i \leq 45^{\circ}$ and a white dwarf mass 
$\rm M_{wd} = (0.6-1.0) M_{\odot}$, corresponds to a  
secondary mass  $\rm M_{s} \geq (0.27-0.37)$ \Msun.

   }

   \maketitle
%

\section{Introduction}
 
The X-ray source \rx\, was discovered from the ROSAT Galactic Plane Survey and identified as a relatively bright (m$_B$ $\sim$ 16) cataclysmic variable (CV) (Motch et al. 1998, hereafter M98). Its X-ray hardness ratios indicate a hard X-ray source, similar to the intermediate polar (IPs) class among cataclysmic variables (CVs). 
IPs are Magnetic Cataclysmic Variables (MCVs), where a white dwarf (WD) with a relatively strong ($\rm B\leq$ 5-10\,MG) magnetic field accretes material from a late-type secondary companion overflowing its Roche-lobe. 
The accretion flow is channelled towards the polar regions where a
strong shock develops above the WD surface (see Patterson (1994) for a review), which means that the signature of channelled accretion in these systems is a strong pulsation at the WD rotational period.
In an effort to elucidate the nature of \rx, we obtained optical observations to search for a signature of the WD rotation and its orbital motion. 

\section{Photometric observations}
\begin{figure}
   \centering
\includegraphics[height=8.5cm,width=8.5cm]{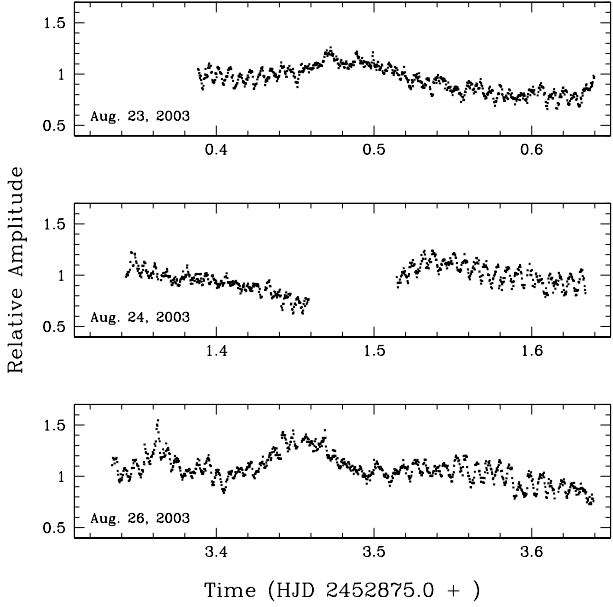}
\caption{\rx\, photometric light curves in relative white light amplitudes with a 30 s resolution. A $\sim$10 min. pulsation is clearly seen on all three nights, superposed on an overall long term variation. }
\label{ph_lc}
\end{figure}

\rx\, was observed from August 23th to 26th, 2003 at the 1.5\,m
Loiano telescope (Bologna, Italy) equipped with the three channel photometer (TTCP) with simultaneous acquisition of target, comparison star, and sky. 
Observations were done without filter with a photometer efficiency peaking around 3800\,\AA\, and a spectral range of $\sim$3000-6000\,\AA\, (Silvotti et al. 2000). The integration time was set to 30\,s in all observations. 
Sky conditions were photometric with a seeing of 1.5\arcsec\, during
all nights, except on August 24 (seeing of 2\arcsec\,) when observations
were also interrupted due to bad sky conditions for $\sim$1.3\,hr.  A diaphragm
of 17\arcsec\, was used for all three channels. \rx\, was estimated at
B$\sim$16\,mag using a nearby comparison star. The log of the observations is reported in Table~\ref{obslog}. The photometric data were reduced using standard procedures including sky
subtraction, extinction correction, and relative photometry. 
The resulting light curve is shown in Fig~\ref{ph_lc}.

\begin{table*}[t!]
\caption{Log of photometric and spectroscopic observations.}
\label{obslog}
\centering
\begin{tabular}{c c c c r r }
\hline \hline
\noalign{\smallskip}
Observations &  Range     & Res. &  Date   & UT(start) & Exposure (min)\\
\noalign{\smallskip}
\hline
\noalign{\smallskip}
Photometry   & White light & 30\,s    & 2003 Aug. 23 & 21:15 & 362  \\
             & White light & 30\,s    & 2003 Aug. 24 & 20:09 & 167  \\
             & White light & 30\,s    & 2003 Aug. 25 & 00:17 & 172  \\
             & White light & 30\,s    & 2003 Aug. 26 & 19:57 & 439  \\
\noalign{\smallskip}
            
Spectroscopy  & 3600-7200 & 5.7\,\AA & 1998 Aug. 24 & 20:35 & 165  \\
              & 3600-7200 & 5.7\,\AA & 1998 Aug. 26 & 20:53 & 210  \\
              & 4300-6000 & 2.8\,\AA & 2004 Jul. 20 & 20:39 & 270  \\
              & 4300-6000 & 2.8\,\AA & 2004 Jul. 21 & 21:08 & 270  \\
              & 4300-6000 & 2.8\,\AA & 2004 Jul. 22 & 20:34 & 300  \\

\noalign{\smallskip}
\hline
\end{tabular}
\end{table*}

The photometric time series were analyzed performing a Discrete 
Fourier Transform (DFT) of the relative amplitudes, after detrending from the low frequency
underlying variations. A significant power was 
present in two groups of frequencies inside the overall range (140-160)\, 
day$^{-1}$ (see Fig.~\ref{ph_fft}). This structure
closely resembles the one observed in the optical power spectra of many IPs with the presence of the spin frequency ($\omega$) and its orbital sideband ($\omega -\Omega$), where $\Omega$ is the orbital frequency. 
The same structure was also found, though with much lower intensity, at twice these
frequencies, which are identified as the first harmonics (see Fig.~\ref{ph_fft}, top panels).  
Given the strong influence of the data window, a number of aliases were also introduced 
that prevent any clear identification of the
true peaks. To remove this effect,  we used the CLEAN algorithm ({\cite{roberts}), 
and results are also shown in Fig.~\ref{ph_fft} (right panels). The most significant peaks were found at
$\omega$ = 151.3$\pm$0.1\,day$^{-1}$ 
with its orbital sideband at $\omega - \Omega$ = 148.0$\pm$0.1\,\,day$^{-1}$
and harmonics at 2$\omega$ = 302.6$\pm$0.1\,day$^{-1}$, where the error bars were determined from the peaks'half widths.
The fundamental and the first harmonic of the spin 
match strictly, while for the beat frequency, the weak peak observed at 
296.6$\pm$0.1\,day$^{-1}$, 
is slightly offset from the expected  beat harmonic ($2\omega - 2\Omega$) frequency. This discrepancy is most likely
due to the very small amplitude (full amplitude 0.8$\%$) and imperfect removal of the data window effect.

   \begin{figure}
   \centering
\includegraphics[height=8.5cm,width=6.1cm,angle=-90]{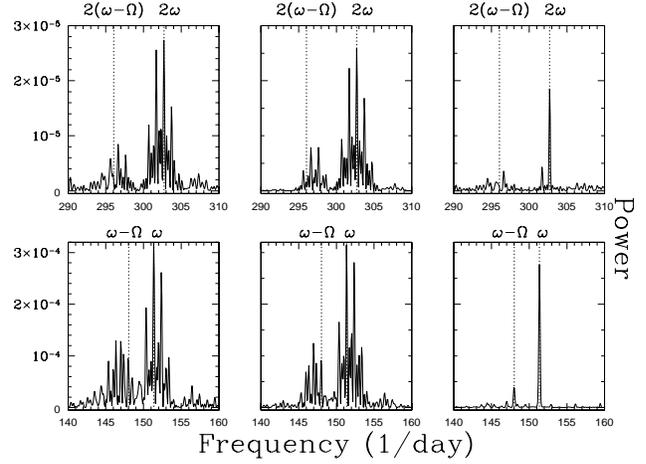}
\caption{Power spectrum of the photometric data (pre-whitened from 
the low frequency trend) around the spin fundamental (bottom) 
and first harmonic (top) frequencies. Note the different power 
scale for the two ranges of frequencies. Left panels: direct 
Fourier transform of the time series, central panels: synthetic 
power spectrum using the best fit periods (see text). 
Right panels: time series "clean" power spectra after removing the 
effect of the data window. The location of the fundamental ($\omega$) 
and orbital sideband ($\omega - \Omega$) are shown by vertical dotted lines. 
For the first harmonics, the small peak around $\sim$ 302\,\,day$^{-1}$ 
is a residual 1-day alias that has been imperfectly removed.}
\label{ph_fft}
\end{figure}

 As the presence of a sideband beat period may significantly affect the period determination, 
the detrended data were fitted with a four frequency sinusoidal
function that includes the fundamental and first harmonic of the spin and beat periods.
The best periods were then found at : 
$\rm P_{\omega}$= 570.8227$\pm$0.013\,s and $\rm P_{\omega - \Omega}$=
583.768$\pm$0.02\,s. 
This fit was used to derive the best spin ephemeris for \rx\, as: 
\begin{equation}
\rm T_{\omega}^{max} = 2\,452876.94785(3) + 0.0066067(2)\,E
\end{equation}
where $\rm T_{\omega}^{max}$ is the heliocentric arrival time of the sine curve maximum.
A quadratic fit on the (O-C) residuals does not yield a significant period derivative 
with an upper limit $\dot{P} \leq$ 2.0 10$^{-6}$ s/s. 

We checked the consistency by producing a synthetic 
DFT of the composite sinusoidal function, including the four most significant 
frequencies above.
Results are shown  in Fig~\ref{ph_fft} (central panels). 
The good agreement between the observed and synthetic spectrum
confirms that the spin and beat modulations have been determined accurately.
The data, pre-whitened from the
low frequency trend, were folded at the derived spin period and the pulse shape 
is shown in Fig~\ref{ph_spin}. The pulse has a full amplitude of $\sim$ 8$\%$ 
and a quasi-sinusoidal shape with a less steep rise than the decay, indicative
of the presence of the first harmonic. 

   \begin{figure}
   \centering
\includegraphics[height=6.cm,width=6.cm]{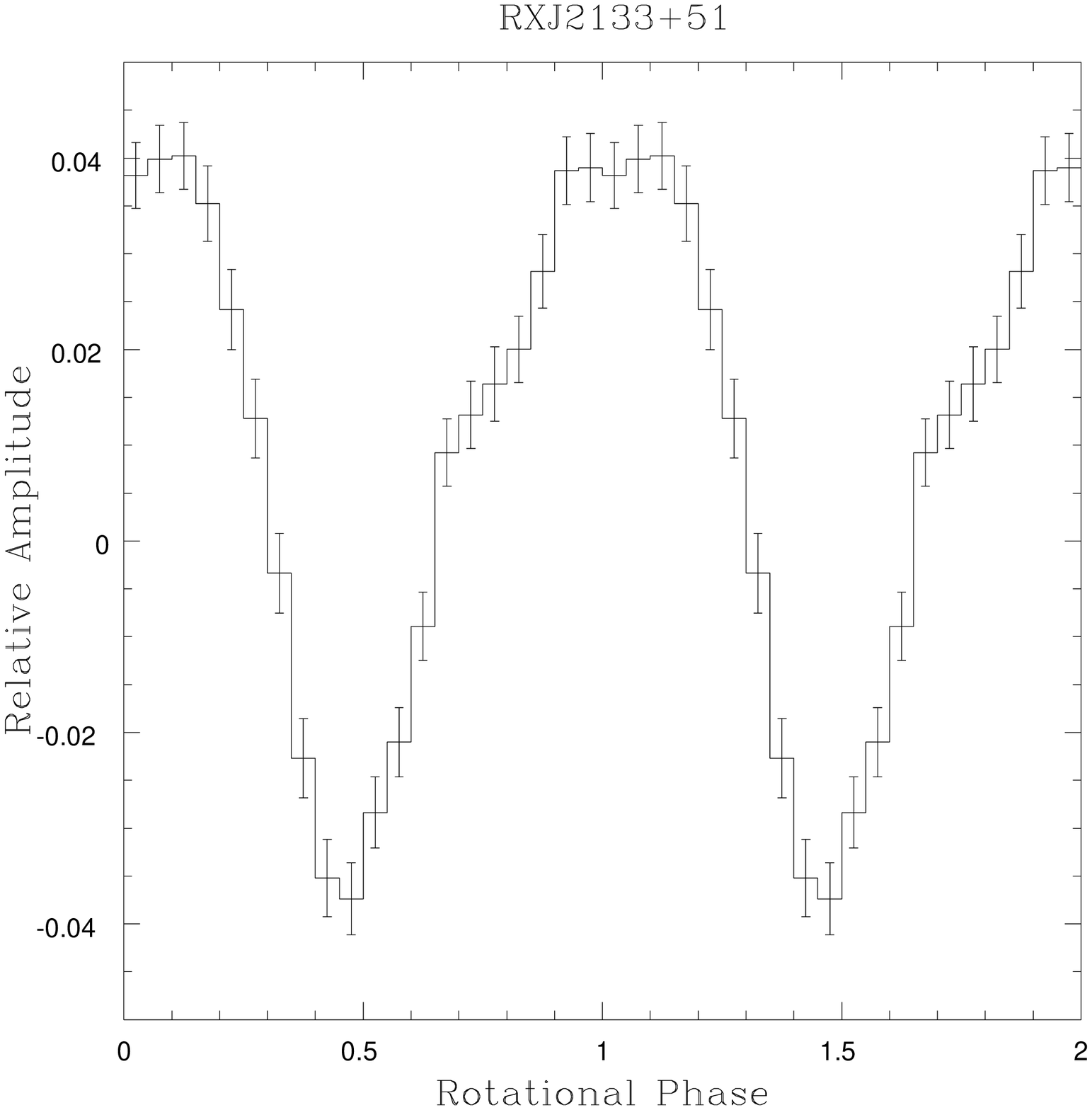}
\caption{Pulse shape of the 571 s spin modulation. A significant distortion 
from a pure sine shape is visible due to the presence of a
first harmonic.}
\label{ph_spin}
\end{figure}

From the photometric data, the measured beat frequency yields an
orbital frequency of $\Omega$ = (3.357$\pm$0.006)\,day$^{-1}$ corresponding to
$\rm P_{\Omega}$ = 7.150$\pm$0.015\,hr, in agreement with the orbital 
period found from spectroscopy (see below). 
The limited number of nights and the probable 
contamination from a likely non-periodic variability do not allow a more precise
determination of the orbital period.

\section{Spectroscopic observations}
\rx\, was observed at the Observatoire de Haute Provence (France)
at two epochs in August 1998 and in July 2004 (see Table~\ref{obslog}). 
Long slit spectra were obtained with the Carelec spectrograph 
(Lemaitre et al. 1990) attached to the Cassegrain focus of the 
193\,cm telescope and using  a CCD EEV (2048x1024 pixels) detector of 
13.5$\mu$m pixel size.
In 1998, the 133\,\AA/mm grating was used, with a slit width of 2\arcsec\,,
leading to a  wavelength coverage of 3600-7200\,\AA\, at an FWHM resolution of
$\sim$ 5.7\,\AA. 
In 2004, spectra were obtained with the 67\,\AA/mm grating and  
a slit of 2\arcsec\,  width, covering the range 4300-6000\,\AA\, at a resolution of 2.8\,\AA.
Exposure times were either 15 min or 30 min in 1998, and 30 min in 2004.
In 1998, clouds were present on the last night (26 Aug). For both nights,
the seeing was greater (3-5$^{"}$) than the slit (2\arcsec\,). 
In 2004, observations were obtained in nearly photometric conditions 
with only thin cirrus during the last night (22 Jul). Seeing was typically of 2.5\arcsec\, 
 to 3\arcsec\,.

Standard reduction was performed in the ESO-MIDAS package, including
cosmic removal, bias subtraction, flat-field correction, and wavelength 
calibration. The wavelength calibration was checked on sky lines that were found within 1.6\,\AA\, and 0.7\,\AA\, from their expected wavelengths in 1998 and 2004, respectively. 
All radial velocity measurements were corrected from the Earth motion and 
from the small instrumental shifts measured on the OI 5577\,\AA\, line
and times were converted in the heliocentric system.
Flux calibration was performed using the standards BD+28 4211 and BD+33 2642.

The mean optical spectrum of \rx\, (Fig.~\ref{sp_mean}) is typical of magnetic CVs with strong emission lines 
of the Balmer series, HeII (4686\AA\, and 5411\AA) and CIII-NIII (4655\AA), superposed on a relatively blue continuum. 
In 2004, when the flux calibration was reliable, the mean B flux was estimated at $\sim 3 \times 10^{-15}$ \ecsa , which corresponds to a B magnitude of $\sim$15.8, similar to what was reported in the identification paper by M98. The characteristics of the main emission lines are given in Table~\ref{lines}. 
Similar equivalent widths were observed in the two observations in 1998 and 2004, but the intensities of the lines change by a factor 2.

\begin{figure}
\centering
\includegraphics[height=8.5cm,width=4.9cm,angle=-90]{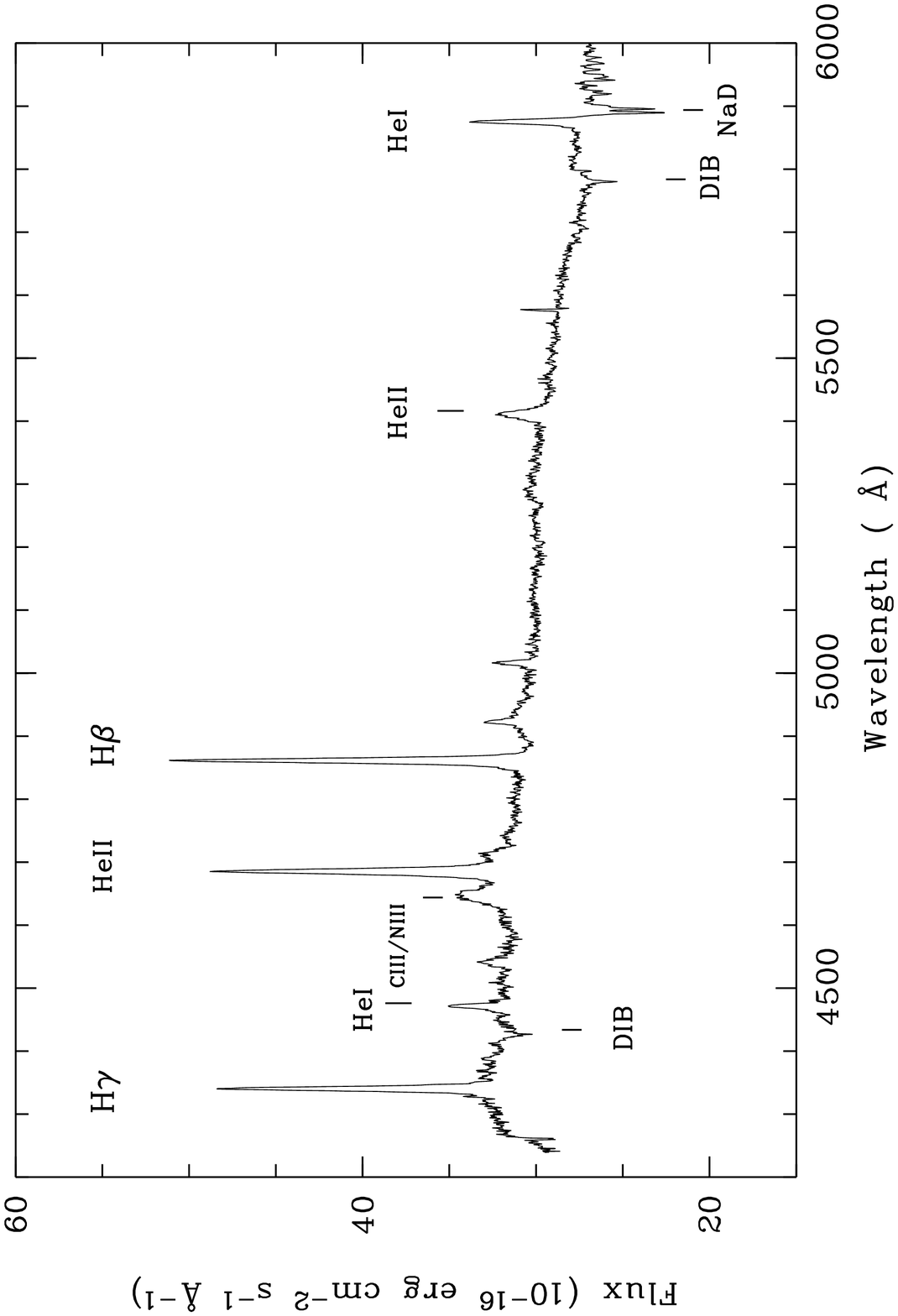}
\caption{The mean optical spectrum of \rx\, obtained in July 2004, showing typical strong  
Balmer and Helium emission lines. Note the significant CIII-NIII and the presence 
of diffuse interstellar bands (DIB) and Na absorption, blended with the HeI (5875\AA) emission.}
\label{sp_mean}
\end{figure}

\begin{table*}
\caption[ ]{Equivalent widths, FWHM and fluxes of the strongest lines}
\label{lines}
\begin{flushleft}
\begin{tabular}{lrrrrrr}
\hline \hline
Line & EW(\AA) & FWHM(\AA) &  Flux(*)  & EW(\AA) & FWHM(\AA) &  Flux(*)  \\
\multicolumn{1}{c}{ } &\multicolumn{3}{c}{August 1998} &\multicolumn{3}{c}{July 2004}\\
\hline 
H$_\delta$        & 2.9(3)  & 7.2    &  56   &-         &-       & -    \\
H$_\gamma$        & 3.9(2)  & 7.5    &  73   & 4.2(2)   & 7.3    & 136  \\
HeI 4471          & 1.0(1)  & 7.5    & 17    & 0.9(1)   & 6.2    &  30  \\
CIII-NIII(+)      & 2.0(1)  & 21.3   &  36   & 2.3(2)   & 26.9   &  71  \\
HeII 4686(+)      & 5.9(1)  & 10.6   & 110   & 5.7(2)   & 10.5   & 182  \\
H$_\beta$         & 6.1(3)  &  8.5   & 109   & 6.4(3)   & 8.0    & 201  \\
HeII 5411         & 1.5(3)  & 12.8   & 23    & 1.4(2)   & 15.2   &  42  \\
HeI 5875(+)       &  2(2)   & 8.1    & 32    & 2.0(2)   &  7.9   &  55  \\
H$_\alpha$        & 16.5(4) & 10.7   &  252  & -        & -      & -    \\
\hline
\multicolumn{7}{l}{Errors bars on last digit in parentheses.}\\
\multicolumn{7}{l}{FWHM: corrected for instrumental resolution (typical errors of 0.6\AA{} in 1998 and 0.3\AA{} in 2004)} \\
\multicolumn{7}{l}{* Line flux in units of \ecs are indicative because of marginal photometric conditions} \\
\multicolumn{7}{l}{+ Measurements derived from Gaussian fits.}
\end{tabular}
\end{flushleft}
\end{table*}

The radial velocities were measured for the main lines. They showed  a significant modulation of 
$\sim$ (30-70) \kms amplitudes that were analysed to constrain the orbital period. The exposure of the individual spectra (30 min.) were close to a multiple of the spin period so that no strong influence from the spin modulation was expected, but the period search was significantly affected by the 1-day sampling of the observation. The best period was determined from an $\chi^{2}$ periodogram where the radial velocity measurements were folded at trial periods and tested against a sine modulation. The most accurate determination was obtained for the strongest line (\Hb) in the longest observation of 2004 (Fig~\ref{sp_xhi2}). The minimum $\chi^{2}$ corresponds to a period of
$\rm P_{\Omega}$=7.193$\pm$0.016\,hr, where the error bar is at a 99\% level, computed for two independent parameters.  A secondary minimum, also observed at  $\sim$10.3\, hr, is a 1-day alias that can be excluded at a level better than 3. $10^{-4}$  from an F-test.
A similar value was obtained independently in 1998 by making use of the velocities of the strongest \Ha\, line, but the accuracy on the period value does not allow us to keep track of the phase between the two observations, so the orbital period cannot be refined. The orbital ephemeris for \rx\, is determined as : 
\begin{equation}
\rm T_{\Omega}= 2453208.1428(28) + 0.29973(67)\,E
\end{equation}
where $\rm T_{\Omega}$ is the predicted heliocentric time of the blue-to-red radial velocity transition. 
The radial velocities of the main lines, folded at the best orbital 
ephemeris above, are shown in Fig.~\ref{sp_vcor}, and the parameters of the 
corresponding best sine fits are given in Table~\ref{velocity}. 
Data are from the 2004 observations obtained with the highest spectral 
resolution, except for \Ha, which was only observed in 1998. 
For this last line , the orbital phase was arbitrarily shifted, 
as the ephemeris were not accurate enough to link the 1998 data. 
The Balmer lines showed similar $\sim$ 70 \kms velocity amplitudes,
while the HeII lines displayed lower value with no significant phase shift. 
Inspection of the line intensities, EW, and widths folded at 
the orbital period do not reveal significant modulation, except 
for a very small variation in width for \Hg and \Hb.

   \begin{figure}
   \centering
\includegraphics[height=8.cm,width=6.cm,angle=-90]{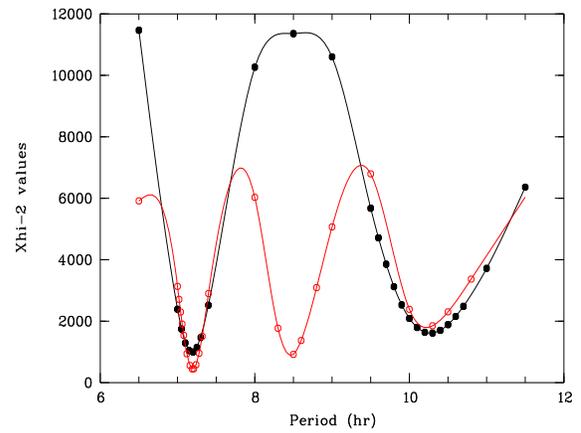}
\caption{$\chi^{2}$ periodogram of the radial velocities of \Hb in 2004 (black line) and \Ha\, in 1998 (grey line). The same best value is found in the two observations for
$\rm P_{\Omega} \sim$ 7.2\,hr. Secondary minima are 1-day (2004) and 2-day (1998) aliases }
\label{sp_xhi2}
\end{figure}

\begin{table}[t!]
\caption{The parameters of the orbital modulation of the emission lines' radial velocities.}
\label{velocity}
\centering
\begin{tabular}{l l l l }
\hline \hline
\noalign{\smallskip}
Lines &  $\gamma$ ( \kms )  & K ( \kms ) &  Phase (*)  \\
\noalign{\smallskip}
\hline
\noalign{\smallskip}
\Hg    & + 12.7 (1.4) & 66.5 (2.0) & 0.003 (0.030)   \\
\Hb    & + 10.6 (0.4) & 67.4 (0.6) & 0.000 (0.009)   \\
\HeII  & - 14.7 (0.9) & 42.2 (1.2) & 0.024 (0.029)   \\
\HeIIb & - 34.2 (5.2) & 33.5 (7.4) & 0.021 (0.220)   \\
\Ha    & - 22.8 (0.4) & 74.8 (0.6) & 0.0   (fixed)   \\           
\noalign{\smallskip}
\hline
\multicolumn{4}{l}{* Phase of the blue-to-red zero crossing }\\
\multicolumn{4}{l}{1-$\sigma$ error bars into parentheses }\\
\end{tabular}
\end{table}

   \begin{figure}
   \centering
\includegraphics[height=12.cm,width=10.cm,angle=-90]{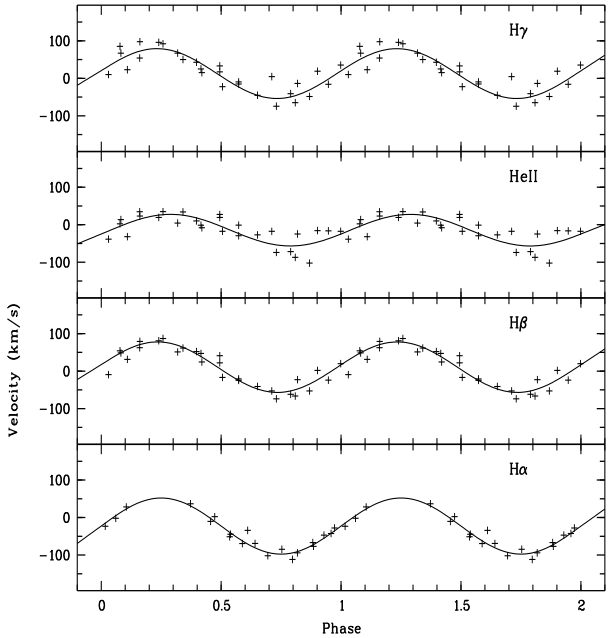}
\caption{Radial velocities of the main lines folded with the orbital period 
$\rm P_{\Omega}$=7.193\,hr.}
\label{sp_vcor}
\end{figure}

\section{Discussion}
The optical characteristics of \rx\, clearly point toward an IP nature of 
the source.
Photometric data show a clear, persistent 571\,s period, which 
we interpret as the spin period of
the accreting white dwarf. The presence of a spin first harmonic
and the detected beat periodicity consistent with a 
$\sim$ 7.2 hr orbital period are features common to many IPs systems,
signaling a probable two-pole accretion with significant reprocessing inside
the binary system (see {\cite{warner}, {\cite{buckley}).
The orbital period is confirmed independently by modulation 
of the emission lines, which yields a more precise value of P = (7.193$\pm$0.016)\,hr.
This is among the longest orbital periods found in intermediate polars as, 
out of $\sim$40 systems, only four 
confirmed IPs (AE\,Aqr, V1062\,Tau, RX\,J1548-4528 and GK\,Per) known so
far have wider orbits ({\cite{norton}).
  
With a longer period than 7\,hr, a secondary of a relatively early type 
(up to K0) can be expected.
Such stars usually show a significant sodium line Na D 
(5889-5896\AA) in absorption. 
This line is indeed detected in the spectrum of \rx\,, partly blended 
with the HeI (5875\AA) line, at the resolution used
in our observations (see Fig.~\ref{sp_mean}). However the Na D line observed 
in \rx\, is at least partly interstellar, 
indicative of both significant distance and absorption. This is confirmed 
by the presence in the spectrum of the diffuse bands at DIB($\lambda$4430) 
and DIB($\lambda$5780) similar to what is observed in some other CVs 
(see {\cite{thorstensen}). 
The equivalent widths of the ($\lambda$4430) and
($\lambda$5780) DIBs are measured at 0.7\AA\, and 0.3\AA\ respectively,  
and are both consistent with a reddenning of $\rm E_{B-V} \sim 0.3$
({\cite{herbig}, {\cite{snow}). If purely interstellar, the
Na D1 ($\lambda$5890) component would have indicated a much higher reddenning
of $\rm E_{B-V} \sim 1.1$ ({\cite{munari}), inconsistent with
the above value and indicating that the secondary also contributes to the line.
From the measured equivalent widths, the star contribution can be estimated to be less than 50 \%.

The Na D equivalent widths were measured on representative stars 
from the {\cite{jacoby} catalogue with spectral types from K0 to M0 
and compared to the EW upper limit measured in \rx. 
If the secondary contributes for less than half of the observed Na D, 
the contribution of a K0, K5, or M0 star to the optical flux is respectively less than 
39\%, 37\%, and 9\%, which translates into a lower limit to their 
distance of 1200, 650, and 600 pc respectively, assuming standard 
absolute magnitudes (Allen 1970). 
X-ray observations of the source by ROSAT do not give better indications 
of absorption and distance.
From the observed ROSAT hardness ratios and the 
measured value HR1=0.90$\pm$0.06 (M98), 
using the HR1-N$_{\rm H}$ distribution for CVs in Fig.4 (M98) gives a
rough estimation of N$_{\rm H}$ $\sim 1 \times 10^{21}$ $\rm cm^{-2}$ 
with, however, a large uncertainty. 
The measured ROSAT flux and 
the count conversion given in M98 yield a [0.1-2.4 keV] luminosity of 
L$_x$ $\sim 1.5 \times 10^{32}$ (d/500pc)$^2$ \es.

A complementary constraint on the system can also be obtained from the 
velocity amplitude of the emission lines, if they are assumed 
to be associated with the white dwarf. The mean velocity of the Balmer 
lines yields a mass function of
f$_m$ = (1.05$\pm$0.21) $\times 10^{-2}$ \Msun. For an inclination 
of 
$i \leq 45^{\circ}$, suggested by the absence of double-peaked lines,
and a white dwarf mass $\rm M_{wd} = (0.6-1.0) M_{\odot}$, , 
the secondary mass is $\rm M_{s} \geq (0.27-0.37) M_{\odot}$, 
consistent with a star at least more massive than M2V. 
 
The relatively long orbital period of \rx\, makes it fall into what has 
been called the "IP gap" ({\cite{schenker}), a lack of IPs into 
the (6-10) hr orbital period interval. From the list of 39 IPs in 
{\cite{norton}, all are either shorter than 6.14 hr or longer 
than 9.87 hr, this last value being the one newly determined 
for 1RXS J1548-4528 by de Martino et al. (2005). Though it could 
still be due to an observational bias, this may indicate a particular 
state of evolution.

With a spin-to-orbit period ratio of 0.02, \rx\, is also a
relatively highly asynchronous system with only 8 IPs showing lower values. 
This suggests a young system still far from equilibrium. 
For magnetic CVs, Norton et al. (2004) have computed the 
equilibrium rate when the spin-up accretion torque is balanced by 
the braking effect of the magnetic torque, using a detailed 
plasma-magnetic field interaction. 
For a value of $\rm P_{\omega}/\rm P_{\Omega}$= 0.02 and an 
orbital period of 7.2hr, the equilibrium is reached for a 
magnetic moment of $\mu \sim 10^{33}$ G.cm$^3$, here assuming a mass ratio $q=0.5$,
but the variation is less than 10\% for q in the range 0.3$\leq$ q $\leq$0.7. 
The magnetic moment is probably much lower in \rx\,, since accretion will only 
take place if the Alven radius (r$_m$) is lower than the corotation radius (r$_{co}$)
(see {\cite{warner}). 
For \rx\,, r$_{co}$ = $(0.87-1.03) \times 10^{10}$ cm (for \Mwd\, in 
the range (0.6-1.0) \Msun). For $\mu \sim 10^{33}$ G.cm$^3$, 
the (r$_m$ $\leq$ r$_{co}$) condition will require an accretion rate that is higher than 
(2.8-3.5) $\times 10^{17}$ g.s$^{-1}$, so much too high for the observed 
X-ray luminosity.
Assuming a more canonical value of $\dot{M} \sim 10^{16}$ g.s$^{-1}$, the magnetic 
moment then has to be lower than (1.2-1.9) $\times 10^{32}$ G.cm$^3$; therefore \rx\, 
is most probably an IP with a weak magnetic field.
A definite confirmation of its IP nature will be possible with 
the detection of strong X-ray pulses expected at the 571\,s period.


\end{document}